\documentclass[aps,prl,superscriptaddress,twocolumn,longbibliography,nofootinbib,notitlepage,preprintnumbers]{revtex4-1}

\usepackage{graphicx}
\usepackage{amsmath}
\usepackage{amssymb}
\usepackage{amsfonts}
\usepackage{amsthm}
\usepackage{hyperref}
\usepackage{color}
\usepackage{soul}
\usepackage{textcomp}
\usepackage{bm}
\usepackage{dsfont}
\usepackage{relsize}
\usepackage{comment}
\usepackage{braket}
\usepackage{enumitem}   
\usepackage{comment}

\usepackage{mathrsfs}
\usepackage{times}
\usepackage{svg}

\theoremstyle{plain}

\numberwithin{obs}{section}

\newcommand{\ketbra}[1]{| #1 \rangle\langle #1 |}

\newcommand{\bigo}[1]{\mathcal{O}\left (#1\right)}
\newcommand{\comments}[1]{}
\newcommand{\ba}{\begin{align}}
\newcommand{\ea}{\end{align}}

\newcommand{\mpl}[1]{\textcolor{black}{#1}}

\begin{document}
\title{Probing coherent quantum thermodynamics using a trapped ion}

\author{O.~Onishchenko$^{*}$}
\affiliation{Institut f\"ur Physik, Universit\"at Mainz, Staudingerweg 7, 55128 Mainz, Germany}

\author{G.~Guarnieri$^{*}$}
\affiliation{Dahlem Center for Complex Quantum Systems, Freie Universit\"{a}t Berlin, 
14195 Berlin, Germany}

\author{P.~Rosillo-Rodes}
\affiliation{Institute for Cross-Disciplinary Physics and Complex Systems, Campus Universitat de les Illes Balears,
E-07122, Palma, Spain}

\author{D.~Pijn}
\author{J.~Hilder}
\author{U.~G. Poschinger}
\affiliation{Institut f\"ur Physik, Universit\"at Mainz, Staudingerweg 7, 55128 Mainz, Germany}

\author{M. Perarnau-Llobet}
\affiliation{Department of Applied Physics, University of Geneva, 1211 Geneva, Switzerland}

\author{J.~Eisert}
\affiliation{Dahlem Center for Complex Quantum Systems, Freie Universit\"{a}t Berlin, 
14195 Berlin, Germany}

\author{F.~Schmidt-Kaler}
\affiliation{Institut f\"ur Physik, Universit\"at Mainz, Staudingerweg 7, 55128 Mainz, Germany}

\begin{abstract}
Quantum thermodynamics is aimed at grasping thermodynamic laws as they apply to thermal machines operating in the deep quantum regime, a regime in which coherences and entanglement are expected to matter. Despite substantial progress, however, it has remained difficult to develop 
thermal machines in which such quantum effects are observed to be of pivotal importance. In this work, we report an experimental measurement of the genuine quantum correction to the classical work fluctuation-dissipation relation (FDR). We employ a single trapped ion qubit, realizing thermalization and coherent drive via laser pulses, to implement a quantum coherent work protocol. The results from a sequence of two-time work measurements display agreement with the recently proven quantum work FDR, violating the classical FDR by more than $10.9$ standard deviations. We furthermore determine that our results are incompatible with any SPAM error-induced correction to the FDR by more than $10\sigma$.
Finally, we show that the quantum correction vanishes in the high-temperature limit, again in agreement with theoretical predictions. 
\end{abstract}
\date{\today}

\maketitle

\def\thefootnote{*}\footnotetext{These authors contributed equally to this work.}\def\thefootnote{\arabic{footnote}}

One of the pillars on which modern physics rests is classical phenomenological thermodynamics. Born out of the scrutiny of the functioning of heat engines in the 19th century, it is one of the most profound physical theories available, offering a wealth of applications within physics and in everyday life. Its strength originates from the fact that while it applies to the description of complex systems, consisting of a large number of microscopic constituents, it largely abstracts from details about these individual physical constituents: It offers a description in terms of a small number of quantities, capturing the relationship between heat, work, temperature, and energy. In recent years, it has become increasingly clear that the rules of thermodynamics must be sharpened for thermal machines operating in the quantum regime, a regime in which coherence, entanglement and quantum fluctuations are expected to 
play a role \cite{Topical,MillenReview,KosloffReview,AndersThermo,LongReview}. 
While such a development was unforeseeable for the founders of thermodynamics, the question of how thermodynamic notions should be altered has become highly relevant in light of experimental progress in quantum engineering. In fact, the past few decades have seen huge leaps towards experimental realizations of meso- and nano-scale devices ~\cite{dubi2011colloquium,blickle2012realization,martinez2016brownian,rossnagel2016single,josefsson2018quantum,peterson2019experimental}, culminating in the emerging quantum technologies \cite{Roadmap}. The tantalising prospect of harnessing quantum features and exploiting them in order to outperform classical counterparts has boosted research across many fields, ranging from quantum biology ~\cite{Huelga_2013}
to quantum computation~\cite{arute2019quantum,SupremacyReview,boixo2018characterizing}.

\begin{figure}[ht]
    \centering
     \includegraphics[width=\columnwidth]{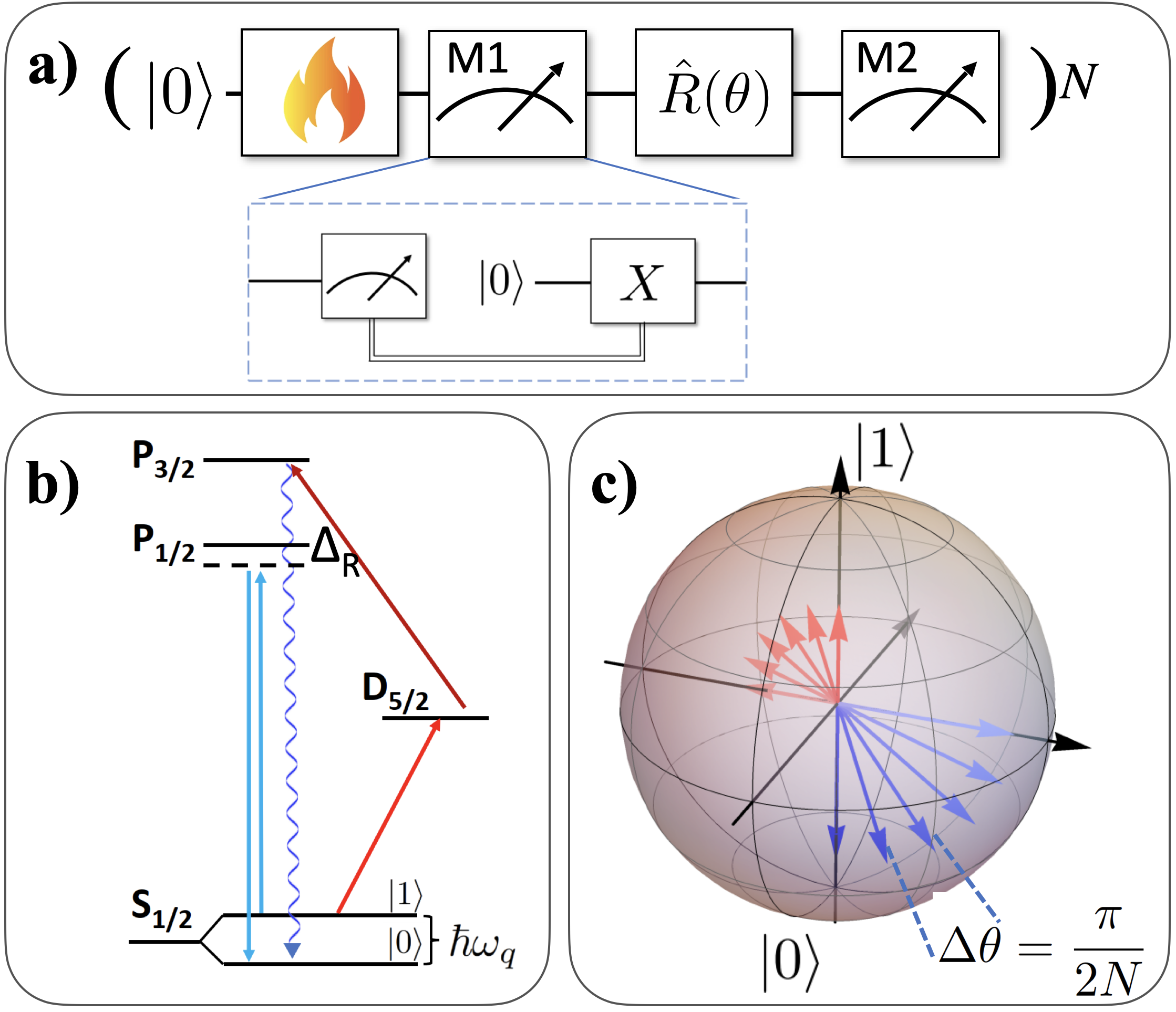}
    \caption{Experimental implementation scheme for detecting quantum work fluctuations using a trapped-ion qubit. \textbf{a)} Sequence of operations for obtaining a single measured value of work. The flame represents thermalization to a Gibbs state in the computational basis. $\ket{0}$ represents qubit initialization. A two-point measurement is realized around a work step, given by a qubit rotation $\hat{R}(\theta)$. The inset shows an non-demolition energy measurement, which
    is emulated via conditional re-initialization. \textbf{b)} The relevant energy levels of a $^{40}\text{Ca}^+$ ion. The qubit is encoded in the Zeeman sub-levels of the $4 ^{2}\text{S}_{1/2}$ ground state. The Raman detuning $\Delta_\text{R}$ is typically 250 GHz.  
    \textbf{c)} Discrete protocol visualized on the Bloch sphere.}
    \label{fig:schematic}
\end{figure}

A rich body of theoretical work has provided guidelines of what to expect in this exciting development
in quantum thermodynamics 
\cite{Topical,MillenReview,KosloffReview,AndersThermo,LongReview,Brandao2015PNAS_secondlaws}. However, to see evidence of \emph{genuine quantum effects} in experimentally realized microscopic thermal machines seems to be harder to come across. 
Thermal machines operating \emph{with} single quantum systems and featuring signatures of quantum coherence have been devised~\cite{Myers_2022,maslennikov2019quantum,Lindenfels2019,Peterson2019,bouton2021quantum,wenniger2022arxiv} \mpl{and compared to their classical counterpart~\cite{Klatzow2019Experimental}}; but the task of actually verifying and statistically significantly observing genuine quantum signatures that have no classical analogue in thermodynamics remains elusive. 
This is in particular relevant when assessing the important question how quantum coherence affects protocols of
actual work extraction.

It has become clear, again from a theoretical perspective, that fluctuations as inspired by the notions of quantum \emph{stochastic thermodynamics}~\cite{Esposito2009review,Campisi2011review,baumer2018fluctuating} 
may be the tool that offers to discriminate quantum from
classical prescriptions and actually provides the answer.
Within this general framework, all thermodynamic quantities (such as work) become stochastic variables at the microscopic level, that can be defined at the level of single trajectories or dynamical realizations~\cite{Esposito2009review,Campisi2011review,Miller2021PRE} and are therefore distributed according to probability distributions from which fluctuations can be computed. Quantum effects most prominently manifest themselves as such fluctuations~\cite{Allahverdyan2014non,Talkner2016}. In particular, these have a prominent impact on the so-called \emph{work fluctuation-dissipation relation} (FDR)~\cite{hermans_simple_1991,Jarzynski1997,speckDistributionWorkIsothermal2004}
\begin{align}
\label{FDRclassical}
\frac{\beta}{2} \mathrm{Var}(W) = \langle W\rangle - \Delta F,
\end{align}
which is valid in linear response, i.e., when the system remains close to thermal equilibrium at all times~\cite{speckDistributionWorkIsothermal2004,miller2019work,scandi2019work}.
Eq.~\eqref{FDRclassical} relates the first two cumulants of the work distribution, i.e., $\langle W\rangle$ and $\mathrm{Var}(W)$, for a slowly driven system in contact with a thermal bath at inverse temperature~$\beta>0$, with $\Delta F$ being  the change in free energy between the two endpoints of the process. Eq.~\eqref{FDRclassical} has moreover been confirmed in various experimental platforms in mesoscopic  systems~\cite{Jun2014,Barker2022}. 

However, it has recently been shown~\cite{miller2019work,scandi2019work} that Eq.~\eqref{FDRclassical}
is violated in the presence of \emph{quantum friction}, i.e., the generation of coherence in the instantaneous energy eigenbasis. The observation of these quantum violations requires two basic ingredients: (i) coherent driving of the system of interest, in the sense that the Hamiltonian at different times does not commute, $[H(t),H(t')]\neq 0$ for some $t'\neq t$, so that coherence is dynamically generated, and (ii) coupling of the system to a thermal bath that drives it into a Gibbs state with respect to the instantaneous Hamiltonian $H(t)$. Trapped ion qubits offer an excellent degree of control in terms of coherent and dissipative operations and thus represent an ideal platform for studying microscopic thermodynamics \cite{An_2014,maslennikov2019quantum,Pijn2022Detecting,Yan2022Verification}.

In this work, we report on the experimental observation of such a \emph{genuine quantum correction} using a single trapped-ion qubit. Our results clearly show a positive quantum correction to the work FDR Eq.~\eqref{FDRclassical} due to quantum coherent driving, which is shown to vanish in the high-temperature limit, where thermal fluctuations dominate. Importantly, we demonstrate that such correction measurements are incompatible by more than $10.9\sigma$ with values obtained from any incoherent (i.e., classical) protocol at finite driving speed and by more than $12.1\sigma$ with values stemming from \emph{state preparation and measurement} (SPAM) errors, thus statistically certifying the genuine quantum nature of our findings.

{\it Experimental implementation.} We encode a qubit in the spin of the valence electron of a $^{40}\text{Ca}^+$ ion ~\cite{Poschinger2009} confined in a segmented Paul trap~\cite{Schulz2008}. Fig.~\ref{fig:schematic} \textbf{b)} shows the relevant energy levels and transitions. A static magnetic field gives rise to an energy splitting of $\omega_q \approx 2\pi \times 10.5~\text{MHz}$ between the Zeeman sub-levels of the $4 ^2\text{S}_{1/2}$ ground state, which are taken to be the logical basis vectors $\ket{0} := \ket{m_j = +1/2}$ and $\ket{1} := \ket{m_j = -1/2}$ ~\cite{Ruster2016}. Qubit initialization to state vector $\ket{0}$ is realized via optical pumping on the  $4 ^2\text{S}_{1/2} \leftrightarrow 4 ^2\text{P}_{1/2}$ transition using a circularly polarized driving field. A state preparation fidelity of better than ~99\% is obtained by an additional pumping stage, consisting of repeated selective depletion of one of the qubit levels via a $\pi$-pulse driving the narrow $4 ^2\text{S}_{1/2} \leftrightarrow 3 ^2\text{D}_{5/2}$ electric quadrupole transition at about 729~nm, followed by depletion of the meta-stable state by driving the $3 ^2\text{D}_{5/2} \leftrightarrow 4 ^2\text{P}_{3/2}$ transition at about 854~nm.

A (thermal) Gibbs state in the logical basis $\{\ket{0},
\ket{1}\}$ is prepared by controlled \textit{incomplete} optical pumping. We perform the pumping scheme on the quadrupole transition as mentioned above, selectively depleting $\ket{0}$. The control over the transfer probability from state vector $\ket{0}$ via the pulse area enables us to prepare any Gibbs state of the form
\begin{align}
\label{eq:gibbs}
\hat{\rho} = \frac{1}{\mathcal{Z}}\left( \ketbra{0} + e^{-\beta \hbar \omega_q} \ketbra{1} 
\right),
\end{align}
where the partition function
$\mathcal{Z}$
ensures normalization.
The effective inverse temperature $\beta>0$ is controlled by the initial excited state probability $p$ via~\cite{Lindenfels2019}
\begin{equation}
    \text{Tr}( \hat{\sigma}_z\hat{\rho}) = 1-2p = \mathrm{tanh}\left(\frac{\beta \hbar \omega_q}{2}\right).
\label{eq:thermalpop}
\end{equation}
We calibrate $\beta$ by repeatedly performing incomplete optical pumping and projective measurement in the logical basis.

Coherent qubit rotations are performed via stimulated Raman transitions driven by a pair of off-resonant beams, far red-detuned from the $4 ^2\text{S}_{1/2} \leftrightarrow 4 ^2\text{P}_{1/2}$ transition, see Fig.~\ref{fig:schematic}b. The  large detuning ensures that the qubit rotations is performed without significant parasitic dissipation. The drive Hamiltonian $\hat{H}_d=\hbar\Omega \hat{\sigma}_x/2$ generates the evolution $\hat{R}(\theta)=\exp(-i (\theta/2) \hat{\sigma}_x)$, where the ``pulse area'' $\theta$ is accurately controlled with acousto-optic modulators. Starting from a logical basis state, the state after applying $\hat{U}(\theta)$ is an eigenstate of the Hamiltonian
\begin{equation}
\label{eq:controlHam}
\hat{H}_\text{coh}(\theta) = \frac{\hbar\omega_q}{2}\left(\sin(\theta)\hat{\sigma}_y-\cos(\theta)\hat{\sigma}_z\right).
\end{equation}
The operation $\hat{R}(\theta)$ creates coherence in the logical basis, such that the Gibbs state \eqref{eq:gibbs} can be prepared in an arbitrary basis. In the following, we set $\hbar\omega_0 := 1$ without loss of generality.
Qubit readout is performed by selective population transfer from $\ket{0}$ to the meta-stable state $3 ^2\text{D}_{5/2}$~\cite{Poschinger2009}. After that, the detection of state-dependent fluorescence using 397-nm light reveals the result $\ket{1}$ (``bright") or $\ket{0}$ (``dark"). This qubit readout is destructive, as the post-measurement state ends up being completely depolarized. Realizing a projection-valued measurement therefore requires re-initializing the qubit after a measurement by optical pumping, followed by a $\pi$-pulse conditional on the previous measurement result, see Fig. \ref{fig:schematic}a. 

\emph{Quantum fluctuation-dissipation relation protocol.} We realize a discrete protocol of alternating driving and thermalization steps as proposed in Ref.~\cite{scandi2019work}, which has the additional benefit of not requiring any dynamical detail about the relaxation process.
Throughout the process, the angle $\theta$ in \eqref{eq:controlHam} is varied from $\theta=0$ to $\theta=\pi/2$ in discrete steps $j=0,\dots,  N-1$, changing the effective Hamiltonian from $\hat{H}^{i}_\text{coh} = -{\hbar\omega_q}\hat{\sigma_z}/2$ to $\hat{H}^{f}_\text{coh} = \hbar\omega_q\hat{\sigma}_y/2$.
For each step, the following operations are carried out: (i)~The qubit is prepared in a Gibbs state at inverse temperature $\beta$, followed by a readout in the computational basis, which yields the outcome $e_{j}\in \{0,1 \}$, (ii)~the qubit is prepared in the state vector $\ket{0}$, followed by a $\pi$-pulse only if $e_j = 1$, (iii)~a coherent rotation $\hat{R}(\Delta\theta)$ by angle $\Delta\theta=\pi/2N$ is implemented, and finally (iv)~a readout, which gives $e_{j}' \in \{0,1 \}$. We consider the difference of these measurements as \textit{work} after step $j$: $w_j=e_{j}' - e_j$; which follows the standard \emph{two-point-measurement} 
(TPM) scheme ~\cite{Talkner2007c,Campisi2009a,Wang2022arxiv}, based on projective measurements of the qubit energy at the beginning and at the end of each driving step. The full protocol is schematically illustrated in Fig.~\ref{fig:schematic}.

The rotation angle $\Delta\theta$ is calibrated experimentally by preparing the qubit in the state vector $\ket{0}$, then making a pulse of length $t_\text{Raman}$ with the Raman beam pair, and finally 
performing a measurement. This experiment is repeated a few thousand times for different $t_\text{Raman}$, while keeping the intensity, and thus the two-photon Rabi frequency $\Omega$ constant. The probability of reading out ``bright" (corresponding to 1) is given by $P_\text{bright} = \sin^2(\Omega t_\text{Raman})/2)$ \cite{Foot2005atomicphysics}, and that can be related to the probability of detecting ``bright" after a rotation on the Bloch sphere by angle $\theta$ from the pole, $P_\text{bright} = \sin^2(\theta/2)$. Even though $P_\text{bright}$ is not a linear function $t_\text{Raman}$, we 
have chosen closely-spaced values of $t_\text{Raman}$ so that $P_\text{bright}$ is linearized. We can then determine the required $t_\text{Raman}$ by linear regression.

For each step, the probabilities to measure positive or negative work within the TPM are
given by $P(w_j=+1)=(1-p)\;\sin^2(\Delta\theta/2)$, $P(w_j=-1)=p\;\sin^2(\Delta\theta/2)$, and $P(w_j=0)=1 - P(w_j=+1) - P(w_j=-1)$, with the initial thermal excited state population $p$ from Eq.~(\ref{eq:thermalpop}). Since each thermalization step resets any information that has been available in the previous state, the protocol is effectively Markovian and the step work probabilities $w_j$ are statistically identical and independent. This justifies starting each step in $\ket{0}$ for the sake of experimental simplicity. The total work accumulated throughout the process is the sum of individual $w_j$: $W= \sum_{j=1}^N w_j$. Due to thermal and eventual quantum fluctuations, $W$ is a stochastic quantity. 

From the first and second moments, i.e., its mean $\langle W \rangle $ and its variance ${\rm Var}(W)$, we quantify violations of the work FDR Eq.~(\ref{FDRclassical}) by introducing the quantity
\begin{equation}\label{QFDR}
\mathcal{Q}=\frac{\beta}{2}{\rm Var}(W) - \left[\langle W \rangle - \Delta F \right].
\end{equation}
Note that for the processes considered, the free energy change $\Delta F$ in 
Eq.~(\ref{FDRclassical}) is invariant under a basis change given by the effective Hamiltonian
Eq.~(\ref{eq:controlHam}), and therefore $\Delta F=0$ and the total work $W$ equals the dissipated work. We also recall that the regime of validity of the work FDR~(\ref{FDRclassical}) is the linear-response or slow driving regime; for the protocols considered here, this means focusing on protocols with $N\gg 1$ and keeping the magnitudes of interest at order~$\bigo{N^{-1}}$~\cite{scandi2019work}. Explicit analytical expressions for the quantities appearing in Eq.~\eqref{QFDR} are provided in the Supplementary Material. 

\begin{figure}[h!]
	\begin{center}
	\includegraphics[width=0.9\columnwidth]{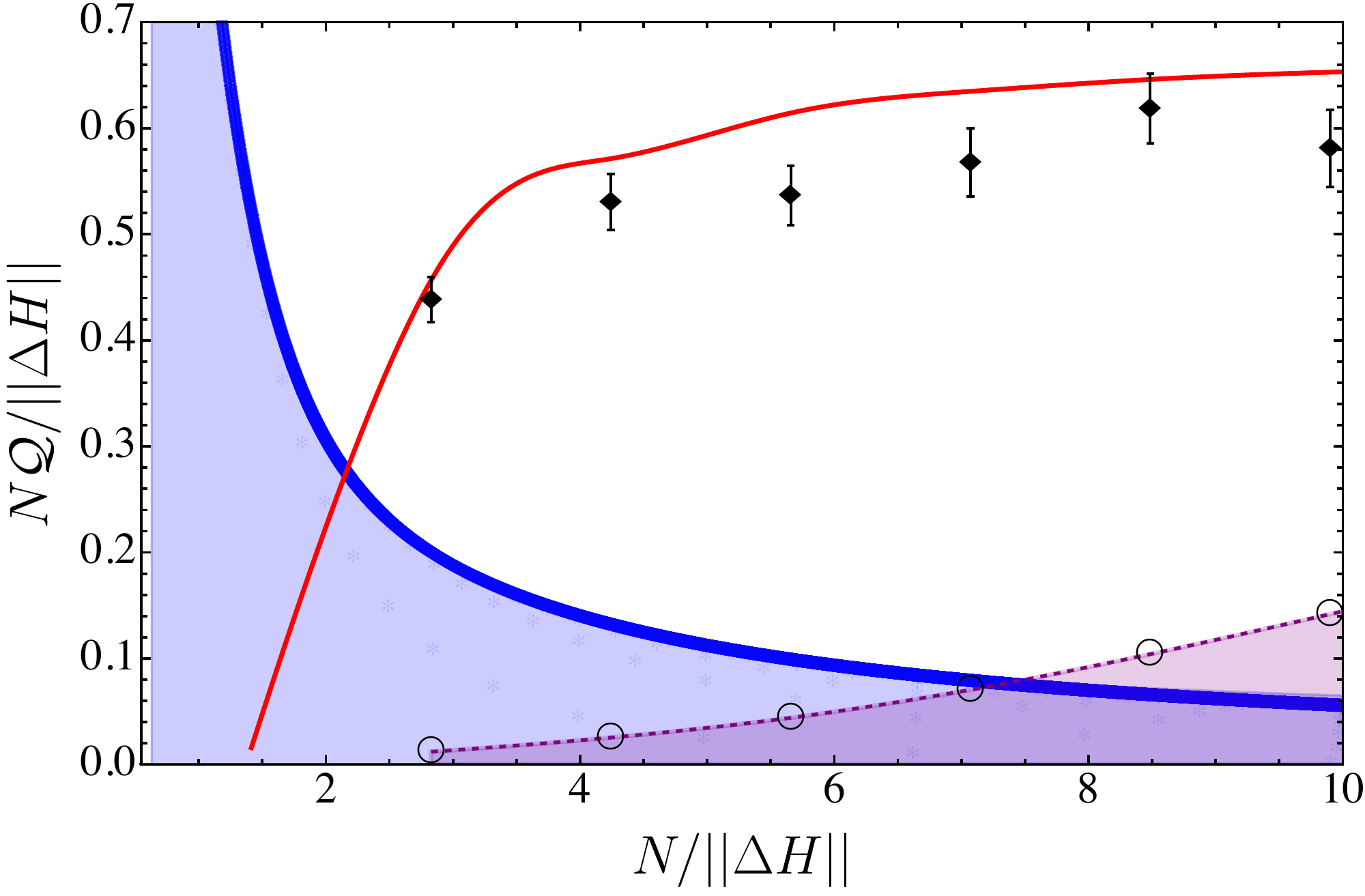}
	\end{center}
	\caption{Quantum correction $N\mathcal{Q}/\|\Delta H\|$ as a function of the inverse of the process velocity $v^{-1} = N/\|\Delta H\|$. The slow driving regime $v\ll 1$ corresponds to the large number of subdivisions $N \gg 1$. The results of the experimental measurements (black dots) are shown together with the 
	theoretical curve (solid red line) calculated for a discrete fully coherent protocol. The plot also displays the simulations of thousands of incoherent processes (blue markers) obtained by both varying all possible qubit's energy gap $\delta\omega = \omega_f-\omega_0 = 2\|\Delta H\|$ and the number of subdivisions $N$ and such that their ratio, i.e., the speed $v=\|\Delta H\|/N$ is within the range plotted. Finally, the values of $N\mathcal{Q}_{\text{SPAM}}/\|\Delta H\|$ are also displayed (black empty circle marker). This plot clearly shows that the experimental points have a finite separation from the region (in light-blue color) of incoherent processes and from the SPAM error region (in light-purple color), thus quantitatively proving that the measured quantum correction $N\mathcal{Q}/\|\Delta H\|$ is a genuine quantum signature.}
	\label{fig:ProofGenuineQuantum}
\end{figure}

\paragraph{Results and discussion.} We first characterize the quantum correction $\mathcal{Q}$ to the FDR by collecting work samples for different values of the step number $N$ at a fixed inverse temperature $\beta=3.413\pm 0.025$ (in units of $\hbar\omega_0$), corresponding to a excited state population $p=0.032\pm 0.001$. For each value of $N$, we repeat the protocol 8000 times and compute the sample mean and sample variances from the total work values $W$ pertaining to each work sample. 
This allows us to reveal the quantum correction Eq.~\eqref{QFDR} for different values of $N$, see Fig.~\ref{fig:ProofGenuineQuantum}.

To quantitatively certify that this value of $\mathcal{Q}$ is a genuine quantum effect and not one caused by thermal fluctuations, finite $N$, or experimental imperfections, we have 
performed an in-depth two-step analysis. 
First of all, it is important to notice that a deviation from the standard FDR Eq.~\eqref{FDRclassical} is theoretically demonstrated to vanish for incoherent processes (i.e., ones where only the eigenvalues of the Hamiltonian change over time, but not the eigenbasis) only \textit{within the slow-driving regime}. This means that, for finite values of the driving speed, one can observe a positive deviation from the classical FDR even for incoherent processes, which would not be a quantum signature. Such incoherent-stemming FDR deviations, however, would decay as 
$N\mathcal{Q} \sim 1/N$, in contrast to those of quantum origin that would instead reach an asymptotic positive value 
$N\mathcal{Q} \sim 1$, see Fig.~\ref{fig:ProofGenuineQuantum}. To rule this possibility out on a quantitative basis, and therefore prove that our observed values of $\mathcal{Q}$ are truly due to the coherent rotations of the qubit's Hamiltonian, we first introduce the notion of a
\emph{speed} as $v = 
{\|\Delta H\|}/{N}$, where $\|\Delta H\|$ denotes the operator norm of the change in the system's Hamiltonian and where the slow driving regime is recovered when $N \gg \hbar\omega_0 (=1)$. 
We then proceed by simulating hundreds of thousands of incoherent 
protocols, i.e., 
such that the Hamiltonian is driven in time from~$\hat{H}_\text{incoh}^i = {-\hbar\omega_0}
\hat{\sigma}_z/2$ to 
$\hat{H}_\text{incoh}^f = {-\hbar\omega_f}\hat{\sigma}_z/2$, thus changing in $N$ discrete steps the qubit's energy gap~$\hbar\omega_j = \hbar\omega_0 + (\omega_f-\omega_0) j /N$, while keeping the energy eigenbasis fixed to~$\hat{\sigma}_z$. It is immediate to see that, for this class of protocols, $\|\Delta H\| = (\omega_f-\omega_0)/2$, while for the coherent protocol performed in the experiment, one has $\|\Delta H\|=1/\sqrt{2}$. In order to quantitatively compare our experimental observations with the values of FDR corrections compatible with incoherent processes for any possible driving speed, we plot the rescaled quantity $N\mathcal{Q}/\|\Delta H\| \equiv \mathcal{Q}/v$.
The results of these simulations are displayed with blue markers in Fig.~\ref{fig:ProofGenuineQuantum}, together with the experimentally measured values of $N\mathcal{Q}/\|\Delta H\|$ (black dots) and the corresponding fully coherent driving protocol (red solid curve). This analysis clearly shows that our experimental points lie beyond the region of values for $N\mathcal{Q}/\|\Delta H\|$ attainable by any incoherent process (light blue shaded region) and therefore provide a striking evidence that those measured corrections are \textit{only} compatible with a genuinely quantum coherent process.
\begin{table}
\begin{tabular}{||c c c c||} 
 \hline
 Point & $\sigma^{\text{stat}}$ & $\delta^{\text{inc}}$ & $\delta^{\text{SPAM}}$ \\ [0.5ex] 
 \hline\hline
 $(2.828, 0.438)$ & $\sigma_1= 0.021$ & $11\sigma_1$ & $ 19.8\sigma_1$\\
 \hline
 $(4.243, 0.530)$ & $\sigma_2= 0.027$ & $14.8\sigma_2$ & $ 18.9\sigma_2$\\
 \hline
 $(5.657, 0.537)$ & $\sigma_3= 0.028$ & $15.5\sigma_3$ & $ 17.5\sigma_3$\\
 \hline
 $(7.071, 0.568)$ & $\sigma_4= 0.032$ & $15.2\sigma_3$ & $ 15.5\sigma_3$\\
 \hline
 $(8.845, 0.619)$ & $\sigma_5= 0.033$ & $16.8\sigma_3$ & $ 15.7\sigma_3$\\
 \hline
 $(9.899, 0.581)$ & $\sigma_6= 0.036$ & $14.4\sigma_3$ & $ 12.1\sigma_3$\\
 \hline
\end{tabular}
\caption{\label{tablesigmas} Verification of the quantum fluctuation dissipation relation. The statistical margin $\sigma^{\text{stat}}$, the distance to the classical prediction from the 
incoherent process $\delta^{\text{inc}}$ and to that due to SPAM $\delta^{\text{SPAM}}$ is given.}
\end{table}

Furthermore, in contrast to the ideal case where energy measurements in the TPM scheme are error-free, experimental measurement-readout errors may occur. In our setup, the second measurement of each TPM has a small but non-zero conditional probability $p_{d\vert 1}$ of incorrectly reading out the qubit as ``dark" when it was in the `bright" state $\ket{1}$, and vice-versa. 
As we show in detail in the Supplementary Material, the introduction of these measurement readout errors leads to a spurious non-zero correction $N\mathcal{Q}_{\text{SPAM}}/\|\Delta H\|$.
As trapped-ion qubits feature low SPAM, we can show that the quantum correction is above $12.1\sigma$ away from the spurious broadening of the FDR due to SPAM for all the experimentally measured points.
We, moreover, show that, at a striking difference from both the quantum coherent $N\mathcal{Q}/\|\Delta H\|$ (which 
reaches an asymptotic constant value for increasing $N$) 
and the fully incoherent correction (which decreases to 
zero as $1/N$), $N\mathcal{Q}_{\text{SPAM}}/\|\Delta H\|$ \textit{linearly increases} with the number of steps $N$. 
This has the important consequence that, in presence of such readout errors, any observation of the genuinely quantum correction Eq.~\eqref{QFDR} would be completely hindered in the quasi-static limit $N \rightarrow \infty$.
It is precisely the outstanding control offered by ultra-cold trapped ion platforms that allowed us to keep track and have very small readout error probabilities, thus allowing to witness a quantum correction to the classical FDR that was statistically shown to be incompatible with both incoherent (i.e., classical) processes \textit{and} also with any value due to the above-mentioned SPAM error.
The black empty circles and the underlying purple region in Fig.~\ref{fig:ProofGenuineQuantum} show the region of values of $N\mathcal{Q}/\|\Delta H\|$ that would be indistinguishable from a worst-case scenario $N\mathcal{Q}_{\text{SPAM}}/\|\Delta H\|$ computed by performing energy measurements without any in-between qubit rotations.
In particular, our result shows that the experimental points have a statistical distance above $12.1\sigma$ both to the region of values of $N\mathcal{Q}$ compatible with incoherent processes and with SPAM errors (see Table ~\ref{tablesigmas} for all the specific values).

\begin{figure}[h!]
	\centering
	\includegraphics[width=.89\columnwidth]{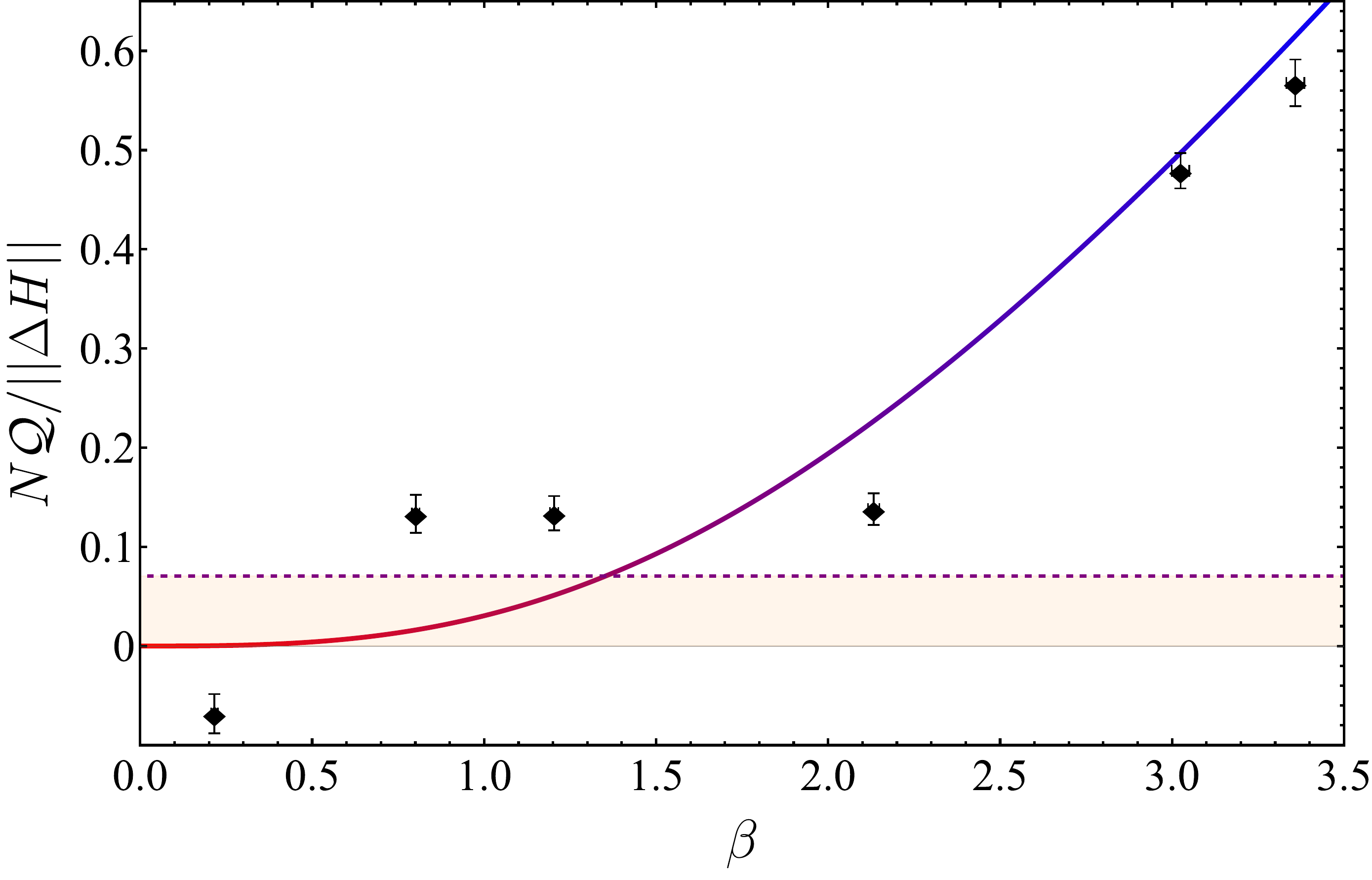}
	\caption{Measured quantum correction $N\mathcal{Q}/\|\Delta H\|$ as a function of the inverse temperature for $N=5$. Error bars are due to the counting statistics with 8000 repetitions. The theory prediction Eq.~\eqref{eq:expansionsNQbeta} is plotted from high (red) to low (blue) temperatures, without free parameters. The maximum SPAM-induced fluctuation readout is denoted by the dashed line.}
	\label{fig:N5vsT}
\end{figure}
We complete our analysis by measuring $N \mathcal{Q}$ as a function of the temperature for a fixed number of subdivisions $N=5$, see Fig.~\ref{fig:N5vsT}. 
First of all, it can be clearly seen that the quantum correction $N \mathcal{Q}$ correctly reproduces the behaviour
\begin{align}\label{eq:expansionsNQbeta}
N \mathcal{Q}/\|\Delta H\| &= (\pi^2\sqrt{2}/4)\left[\beta/2  - \tanh\left(\beta/2\right) \right] +\mathcal{O}(1/N) \notag\\
&\simeq (\pi^2\sqrt{2}/4) \left[0+ \mathcal{O}(\beta^2)\right]+\mathcal{O}(1/N)
\end{align}
where the second line shows a quadratic decrease to zero in the high-temperature limit, where thermal fluctuations dominate (see Supplementary Material for details). Instead, at low temperatures, the excess fluctuations $\mathcal{Q}$ emerge from quantum coherence, which also leads to a non-Gaussian work distribution~\cite{scandi2019work}. 

\paragraph{Conclusions and outlook.} 
In this work, we have exploited the controllability of a trapped-ion qubit platform to perform a measurement of a  quantum thermodynamic signature, namely the quantum correction to the work fluctuation-dissipation relation. 
This has been realized by performing a sequence of $N$ alternating coherent drives, thermalization steps and energy non-demolition measurements on a single qubit.
Our result has revealed a quantum correction to the classical work FDR Eq.~\eqref{FDRclassical} which was proven to be statistically incompatible by more than $10.9$ standard deviations with any incoherent protocol, and incompatible by more than $12.1$ standard deviations with any SPAM-induced error. This conclusion thus certifies the genuine quantum nature of our measurements. 
Moreover, we have shown that the spurious correction to the FDR induced by small, but non-zero, error probabilities in the measurement readout, which we called $N\mathcal{Q}_{\text{SPAM}}/\|\Delta H\|$, has the general property of linearly growing with the number of subdivisions $N$, a scaling in stark contrast both with the incoherent $N\mathcal{Q}_{\text{inco}}/\|\Delta H\|$, which decreases as $~1/N$, and with the quantum correction $N\mathcal{Q}/\|\Delta H\|$, which approaches a constant positive asymptote. 

We believe that our research represents a significant result in the direction of experimental observation and certification of genuine quantum effects in small-scale platforms, thus pushing forward the still painfully scarce body of literature reporting first experimentally observed quantum effects in quantum thermodynamics, relating to squeezed resources \cite{maslennikov2019quantum}, quantum effects in spin quantum heat engines \cite{Peterson2019}, steps towards quantum heat engines driven by atomic collisions \cite{bouton2021quantum}, coherence between solid-state qubits and light fields \cite{wenniger2022arxiv}, signatures of internal coherence \cite{Klatzow2019Experimental}, or verifications of quantum fluctuation relations \cite{LutzExpBayesian}.
Furthermore, our findings tie in with recent insights into
\emph{coherence as a resource} \cite{PhysRevLett.119.140402}, which in our case resulted in measurable quantum 
signatures in the work statistics.
Finally, this invites further exciting endeavours: for example, 
observing quantum effects in quantum field thermal machines in the realm of \emph{quantum many-body physics}~\cite{PRXQuantum.2.030310}. It also 
seems conceivable that further quantum corrections can be measured in work extraction experiments, witnessing temporal coherence reflecting highly \emph{non-Markovian quantum dynamics} \cite{Markov, RevModPhys.88.021002, SusanaNoMarkov}, a feature that could be exploited in order to further further curb down the SPAM measurement-readout errors, allowing to access slower protocols at higher $N$.
One may also bring these experimental results into contact with a body of theoretical work on \emph{single-shot work extraction} \cite{Brandao2015PNAS_secondlaws, WorkWilming, PhysRevX.6.041017}.
It is the hope that the present work will
stimulate further efforts of experimentally exploring the deep quantum regime in quantum thermodynamics. 

\paragraph{Acknowledgements.} We warmly thank J.~Anders for 
helpful comments, and H.~Miller and M.~Scandi for insightful discussions. 
This work has been funded by the DFG (FOR 2724, for which this is an inter-node collaboration reaching an important milestone, and CRC 183) and the FQXI. 
G.~G.~acknowledges fundings from European Union's Horizon 2020 research and innovation programme under the Marie 
Sklodowska-Curie Grant Agreement INTREPID, No.~101026667.
MPL acknowledges funding from
Swiss National Science Foundation through an
Ambizione Grant No. PZ00P2-186067.


\widetext
\clearpage
\begin{center}
\textbf{\large Supplementary Material 
}\\
\end{center}
\setcounter{equation}{0}
\setcounter{figure}{0}
\setcounter{table}{0}
\setcounter{page}{1}
\makeatletter
\renewcommand{\theequation}{S\arabic{equation}}
\renewcommand{\thefigure}{S\arabic{figure}}

This Supplementary Material contains additional details about the data analysis and the theoretical calculations. In particular, Section A provides further insight and derivation of Eq.~(6) of the main text, while Section B contains a detailed discussion, organized in subsections, of all relevant errors used in the generation of Figs.~2 
and~3 of the main text. 

\section{Section A: Analytic expressions for the cumulants of the work distribution}
\label{sec:AppTheor}

In this Section, we provide explicit expressions for the first and second cumulants of the work distribution distribution, i.e., its mean $\langle W \rangle $ and its variance ${\rm Var}(W)$, calculated for the proposed protocol.
In order to do so, we need to make use of the expressions for the step work probabilities to measure positive or negative work within the TPM protocol, i.e.,
\begin{align}\label{TPM}
&P(w_j=+1)= (1-p)\;p_f ,\nonumber \\
&P(w_j=-1)= p\;p_f ,\nonumber \\
&P(w_j=0)= 1-P(w_j=+1)-P(w_j=-1).
\end{align}
Here, $p$ denotes the thermal excited state population from Eq.~(3) of the main text and $p_f$ denotes the flip probability induced by the qubit rotation.

In the case of the coherent protocol describing the driving $\hat{H}^{i}_\text{coh} = -{\hbar\omega_q}\hat{\sigma_z}/2$ to $\hat{H}^{f}_\text{coh} = \hbar\omega_q\hat{\sigma}_y/2$, the ideal flip probability for the case of $N$ subdivisions is given by
\begin{equation}
p_f^{(\text{ideal})}=\sin^2\left(\frac{\pi}{4 N}\right).
\end{equation}
Using this probability, one can compute the first two cumulants of the total work accumulated, which read
\begin{align}
&\langle W \rangle  = N \left(1-2p\right)  \sin^2\left({\frac{\pi}{4N}}\right),
\nonumber\\
&\text{Var}(W) = N  \sin^2\left({\frac{\pi}{4N}}\right) \left(1 -  \sin^2\left({\frac{\pi}{4N}}\right) (1-2p)^2\right). 
\end{align}
Expanding these expressions at leading order in the number of steps corresponds to the slow-driving regime,
\begin{align}
&\langle W \rangle  = \tanh\left(\frac{\beta}{2}\right)  \frac{\pi^2}{4N}+\mathcal{O}(1/N^2),
\nonumber\\
&\text{Var}(W) =    {\frac{\pi^2}{4N}}+\mathcal{O}(1/N^2). 
\end{align}
This implies that the quantum correction $\mathcal{Q}$, in the protocol we experimentally implement, has the following analytical expression
\begin{align}
    &\mathcal{Q} =\frac{\beta}{2}\text{Var}(W)-\langle W \rangle \notag\\
    &= N  \sin^2\left({\frac{\pi}{4N}}\right) \left[ \frac{\beta}{2}\left(1 -  \sin^2\left({\frac{\pi}{4N}}\right) (1-2p)^2\right) - (1-2p) \right] 
    \label{eq:Qanalytic}
    \\ &= \frac{\pi^2}{4N}\left[\frac{\beta}{2}  - \tanh\left(\frac{\beta}{2}\right) \right] +\mathcal{O}(1/N^2),
\end{align}
where we use the relation $1-2p = \tanh{\beta/2}$ (we remind that we set $\hbar\omega_0 = 1$) .
First of all, this expression clearly shows that $\mathcal{Q} \geq 0$ for all $N$, since
\begin{align}
\frac{\beta}{2}  \geq \tanh\left(\frac{\beta}{2}\right) 
\end{align}
for all 
$\beta>0$.
Furthermore, it highlights that the quantum correction vanishes in the high-temperature limit $\beta \to 0$ as $\tanh(x)\approx x+ \mathcal{O}(x^2)$, thus leading to the standard fluctuation-dissipation relation (see Eq.~(1) of the main text).

\section{Section B: Error estimation}\label{sec:analysis}

We discuss how statistical and systematic errors for the quantity $N\mathcal{Q}/\|\Delta H\|$  shown in Fig.~2 and ~3 of the main text (with $N$ being the number of subdivisions and $\|\Delta H\|$ being the norm of the change in the qubit's Hamiltonian) are estimated.

\subsection{Subsection B.1: Parameter extraction}\label{sec:binomial}

Each measured value of the quantity $N\mathcal{Q}/\|\Delta H\|$ is estimated from $M=~8000$ independent runs, each consisting of $N$ \emph{two-point energy 
measurements} (TPM). Each TPM is comprised of two single-qubit readouts in the computational basis. The inverse effective temperature $\beta$ is determined by the probability $p$ of the qubit to be in $\ket{0}$ after the first measurement, according to Eq.~(3) of the
main text. With a binomial statistical error $\sigma^{\text{(b)}}=\sqrt{p(1-p)/(N\;M)}$, the statistical error of $\beta$ is given by \begin{align}
\sigma_{\beta}=
\frac{d\beta}{dp}\vert_{p}\sigma^{\text{(b)}}.
\end{align}
The probability of the result of the second measurement for each TPM being different from the first result is given by the flip probability $p_f$ and can therefore be estimated from the overall frequency of flipped results. It is also subject to the same binomial statistical errors $\sigma^{\text{(b)}}_{f}=\sqrt{p_f(1-p_f)/(N\;M)}$. 

\subsection{Subsection B.2: Statistical errors of $\mathcal{Q}$}\label{sec:staterr}

The measured values of the quantum correction to the FDR $\mathcal{Q}$ are computed from: $\beta$, the average work $\langle W\rangle$ and its variance $\text{Var}(W)$ (see Eq.~(5) 
of the main 
text). 
The resulting statistical error of $\mathcal{Q}$ is computed via bootstrapping, where for each parameter set $(N,\beta,\theta=\pi/(4 N))$, 200 artificial data sets are generated, each consisting of $N M$ statistically independent TPMs. This is done by randomly initializing the qubit in the ``dark" state $\ket{0}$ with a probability $p$ and then flipping it with a probability $p_f$.
For each of the 200 data sets, the mean work, work variance and $\beta$ are determined, from which a bootstrapping value of $\mathcal{Q}$ is computed. The bootstrapping statistical error $\sigma_{\mathcal{Q}}^{(BS)}$ of a given $\mathcal{Q}$ value is the sample standard deviation of the the 200 $\mathcal{Q}$ values obtained for each data set. 
Finally, $\sigma_{\mathcal{Q}}^{(BS)}$ includes the bootstrapping errors $\sigma_{\beta}^{(BS)}$ on the inverse temperature (through Eq.~\eqref{eq:Qanalytic}), which are shown as horizontal error bars in Fig.~(3) of the main text.

\subsection{Subsection B.3: State preparation and measurement (SPAM) errors}

Both state preparation and measurement readout of the qubit are error-prone. After the first measurement of each TPM, the qubit is reinitialized via optical pumping on the quadrupole transition and an optional $\pi$-pulse. The state preparation error for both these processes is very small $\lesssim$~0.1\%, and therefore it can be neglected.
For each TPM, the state is re-prepared based on the result of the first measurement, and the true value of $p$ is given by the dark event probability of the first measurement slot. The determination of $\beta$ can be considered to be free of state preparation and measurement errors.

However, measurement errors have to be taken into account for the second measurement of each TPM, based on the conditional error probabilities $p_{b\vert 0} ~(p_{d\vert 1})$ of incorrectly reading out qubit as ``bright" (``dark") when it has 
actually been in $\ket{0}(\ket{1})$. These conditional error probabilities are determined from separate calibration measurements, which merely consist of state preparation both in $\ket{0}$ and $\ket{1}$ and readout. Using at least 10000 shots for each calibration measurement, we determine readout error rates of about $0.4\pm 0.1$~\%. 

The measurement error leads to modified probabilities to measure nonzero work events within each TPM protocol. In particular, Eq.~\eqref{TPM} becomes
\begin{eqnarray}\label{TPMSPAM}
P'(w_j=+1) &=& (1-p_{d\vert 1}) P(w_j=+1)+ p_{b\vert 0} P(w=0), \nonumber \\
P'(w_j=-1) &=& (1-p_{b\vert 0}) P(w_j=-1)+ p_{d\vert 1} P(w_j=0),
\end{eqnarray}
where $P(w_j=\pm 1, 0)$ are the probabilities defined in Eq.~\eqref{TPM}.
Eq.~\eqref{TPMSPAM} gives rise to spurious values of the correction to the classical FDR we we call $\mathcal{Q}_{\text{SPAM}}$. Specifically, in order to single out the effect of such SPAM errors, we compute $\mathcal{Q}_{\text{SPAM}}$ when coherent drive on the qubit's Hamiltonian is performed (i.e., $\theta=0$) and only TPM readouts are carried out according to Eq.~\eqref{TPMSPAM}. 
This quantity provides the worst-case estimation of a spurious FDR correction, which reads
\begin{equation}\label{QSPAMan}
   \mathcal{Q}_{\text{SPAM}} = N\left[ \frac{\beta}{2} \left[-(p \, p_{d\vert 1} - (1-p) p_{b\vert 0})^2-p
   p_{b\vert 0}+p \,p_{d\vert 1}+p_{b\vert 0}\right]
   + \left[p\, p_{d\vert 1} - (1-p)
   p_{b\vert 0} \right]\right].
\end{equation}
Eq.~\eqref{QSPAMan} clearly shows that this quantity linearly grows with the number of subdivisions $N$, thus becoming the predominant contribution in the slow driving regime $N \gg 1$. 

\subsection{Subsection B.4: Parameter drifts}

The parameters $\beta$ and $\theta$ are determined by laser intensities, which are subject to drifts over the course of the data acquisition. For each value of $\mathcal{Q}$, the parameters are estimated from $M N$ runs. In order to analyze the impact of such potential drifts, these runs are partitioned into $(M N)/K$ bins of size $K$ each, and the probabilities $p$ (from which $\beta$ is computed) and $p_f$ (from which $\theta$ is computed) are calculated for each bin. 
Then, the standard deviation among the various $p$ and $p_f$ is computed and compared to the standard deviation expected from a binomial distribution, as explained in Subsection B.1. For all data sets, no significant excess spread is observed, i.e., $\delta p,\delta p_f <$~0.01 throughout each estimation of $\mathcal{Q}$. We conclude that drift errors are therefore negligible.

\end{document}